\documentclass[%
 aip,
 amsmath,amssymb,
 reprint,%
]{revtex4-1}

\usepackage{graphicx}% Include figure files
\usepackage{dcolumn}% Align table columns on decimal point
\usepackage{bm}% bold math
\usepackage[utf8]{inputenc}
\usepackage[T1]{fontenc}
\usepackage{mathptmx}
\usepackage{etoolbox}

\usepackage{braket}
\usepackage{xr}

\makeatletter
\def\@email#1#2{%
 \endgroup
 \patchcmd{\titleblock@produce}
  {\frontmatter@RRAPformat}
  {\frontmatter@RRAPformat{\produce@RRAP{*#1\href{mailto:#2}{#2}}}\frontmatter@RRAPformat}
  {}{}
}%
\makeatother
\begin{document}

\def\simlt{\mathrel{\lower .3ex \rlap{$\sim$}\raise .5ex \hbox{$<$}}}

\preprint{AIP/123-QED}

\title[Latched readout for the quantum dot hybrid qubit]{Latched readout for the quantum dot hybrid qubit}
% Force line breaks with \\
\author{J. Corrigan}
\thanks{These authors contributed equally to this work.}
\affiliation{Department of Physics, University of Wisconsin-Madison, Madison, WI 53706, USA}
\author{J. P. Dodson}
\thanks{These authors contributed equally to this work.}
\affiliation{Department of Physics, University of Wisconsin-Madison, Madison, WI 53706, USA}
\author{Brandur Thorgrimsson}
\affiliation{Department of Physics, University of Wisconsin-Madison, Madison, WI 53706, USA}
\author{Samuel F. Neyens}
\affiliation{Department of Physics, University of Wisconsin-Madison, Madison, WI 53706, USA}
\author{\\T. J. Knapp}
\affiliation{Department of Physics, University of Wisconsin-Madison, Madison, WI 53706, USA}
\author{Thomas McJunkin}
\affiliation{Department of Physics, University of Wisconsin-Madison, Madison, WI 53706, USA}
\author{S. N. Coppersmith}
\affiliation{Department of Physics, University of Wisconsin-Madison, Madison, WI 53706, USA}
\affiliation{University of New South Wales, Sydney, NSW 2052, Australia}|
\author{M. A. Eriksson}
\affiliation{Department of Physics, University of Wisconsin-Madison, Madison, WI 53706, USA}
 \email{maeriksson@wisc.edu}

\date{\today}% It is always \today, today,

\begin{abstract}
A primary method of reading out a quantum dot hybrid qubit involves projection of the logical basis onto distinct charge states that are readily detected by an integrated charge sensing dot. However, in the simplest configuration, the excited charge state decays rapidly, making single-shot readout challenging. Here, we demonstrate a readout procedure where the qubit excited state is latched to a metastable charge configuration whose lifetime is tunnel rate limited, persisting here as long as 2.5 ms. Additionally, we show that working in the (4,1)-(3,2) charge configuration enables a latched readout window that is larger and more tunable than typical charge configurations, because the size of the readout window is determined by an orbital splitting rather than a valley splitting.
\end{abstract}

\maketitle

An integral requirement of a system viable for quantum computing is reliable and high-resolution readout of the qubit states \cite{DiVincenzo:2000p1536}. In quantum dot systems, the main sensing mechanism is a charge sensor – another quantum dot (QD) or a quantum point contact that is capacitively coupled to the QD qubits and thereby is sensitive to changes in the electron occupation of the qubit QDs. However, many qubit incarnations are spin-based, like the Loss-Divencenzo\cite{Loss:1998p120, DiVincenzo:2000p1642}, Singlet-Triplet \cite{Petta:2005p2180,Shulman:2012p202,Reed:2016p110402}, and Quantum Dot Hybrid Qubit (QDHQ)\cite{Shi:2012p140503,Kim:2014p70}, such that the integrated charge sensor in the device is not able to directly detect the difference between qubit states. Thus, readout of the qubit requires mapping of the energy levels such that one state allows electrons to tunnel and the other state disallows tunneling. This physics is used in Elzerman readout \cite{Elzerman:2004p431}, Pauli spin blockade (PSB)\cite{Johnson:2005p483,Koppens:2006p766}, and latched readout \cite{Studenikin:2012p233101, Harvey-Collard:2018p021046,Corrigan:2021p127701}. Unlike Elzerman readout and PSB, which are limited in the readout length by T1 decay time of the qubit, latched readout involves transferring the excited qubit state to a metastable state of different charge occupation, the tunneling out of which in theory can be tuned arbitrarily long. This method provides much greater flexibility of the measurement time dynamics, and also maps the qubit states to two different total electron numbers, to which the charge sensor is more sensitive than it is to a polarizing charge transfer between two quantum dots.

In this Letter, we demonstrate a latched readout process where the excited charge state of a QDHQ is mapped to a tunnel-limited metastable charge configuration.  We find this extends the duration of the charge state corresponding to the qubit excited state up to 2.5 ms, more than two orders of magnitude longer than the duration of an unlatched readout state~\cite{Kim:2014p70}. Here we operate a QDHQ in the (4,1)-(3,2) charge configuration, and we demonstrate the use of the (3,1) charge state for latching. We show that the 5-electron QDHQ has the further advantage of enabling a latched readout window that is larger and more tunable than 3-electron configurations in silicon,

\begin{figure}[t!]
\includegraphics[width=0.45\textwidth]{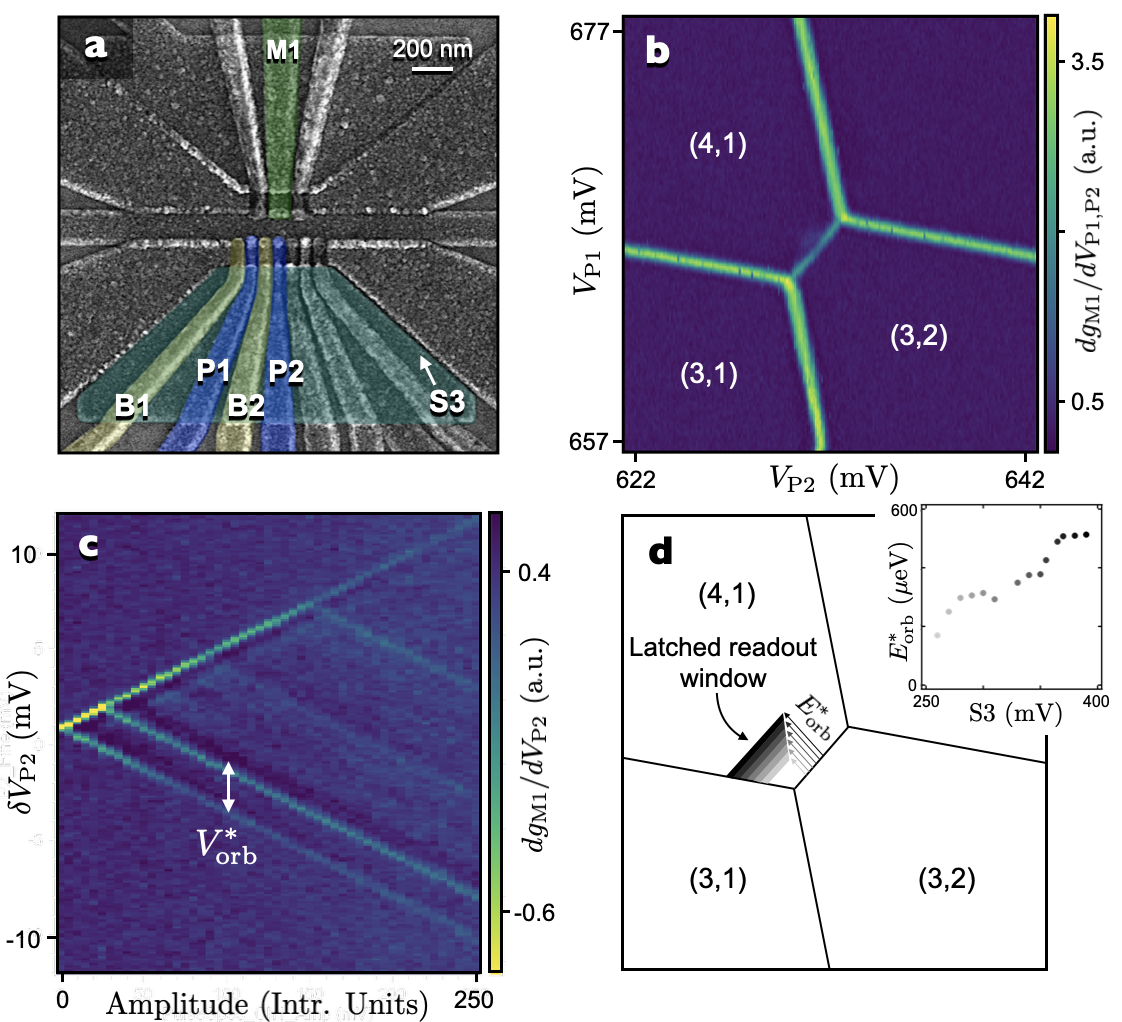}
\caption{\label{fig:fig1} (a) False-colored scanning electron microscope image of a nominally identical device. A double dot is configured beneath plunger gates P1 and P2 (blue) near the (4,1)-(3,2) charge anticrossing. Barrier gates B1 and B2 (yellow) are used to control the tunnel coupling in and out of the P1 and P2 dots. (b) Stability diagram of the (4,1)-(3,2) anti-crossing where the 5-electron quantum dot hybrid qubit (QDHQ) is operated. (c) Measurement of the multi-electron orbital splitting ($E^*_{\text{orb}}$) where electron-electron interactions have suppressed the orbital splitting from the single-electron value. (d) Enhanced readout of the 5-electron QDHQ is enabled by using a latched readout scheme. The latched readout window is shown as the colored triangular regimes. The size of the readout window is determined by the multi-electron orbital splitting, which in the inset is shown to be highly tunable. }
\end{figure} 

\noindent such as (2,1)-(1,2), because the size of the readout window for the 5-electron QDHQ is determined by an orbital splitting, which can be much larger than the typical valley splittings that set the readout window size for the 3-electron QDHQ.

Fig.~\ref{fig:fig1} shows spectroscopic and electrostatic characterization of a 5-electron double quantum dot. Fig.~\ref{fig:fig1}(a) is a scanning electron microscope image of a nominally identical device to the one measured here. The device is fabricated using a three-layer overlapping aluminum gate fabrication process as described in Ref.~\cite{Dodson:2020p505001}. Gates that are integral to the experiment are false-colored. The sensor dot M1 is false-colored in green, the two plunger gates defining the double quantum dot are blue, the barrier gates controlling the tunnel rate between the dots and to the electron reservoirs are shown in yellow, and the screening gate S3 which is used to tune the orbital confinement of the dots is teal.

Figure~\ref{fig:fig1}(b) shows a charge stability diagram of a double quantum dot beneath gates P1 and P2, measured in the ($N_{P1}$, $N_{P2}$) $\rightarrow$ (4,1)-(3,2) regime via lock-in amplifier detection. By working in the (4,1)-(3,2) charge configuration, the first excited state in the P1 dot (containing four electrons) is orbital-like, such that two electrons fill the ground and first excited valley state \cite{Borselli:2011p123118, Lim:2011p335704, Yang:2012p115319}. Fig.~\ref{fig:fig1}(c) shows a pulsed-gate spectroscopy measurement of the multi-electron orbital splitting ($E^*_{\text{orb}}$) in the P2 quantum dot, measured to be between approximately 200--500 $\mu$eV. This energy is significantly larger than the valley splitting in the two-electron regime, measured to be between 25--60 $\mu$eV \cite{Dodson:2021p14702}. Such small valley splittings would make charge-mapped readout difficult by limiting the size of the energy window in which the qubit eigenvalues have different charge signatures. Thus, adding two electrons to one of the quantum dots increases the size of the readout window in devices where the valley splitting is less than the orbital splitting, which is very common.  

The multi-electron orbital splitting is also found to be highly sensitive to changes in the screening gate voltages, enabling tunability of the latched readout window. Fig.~\ref{fig:fig1}(d) shows how the multi-electron orbital splitting relates to the size of the latched readout window. The location of the latched readout window within the charge stability diagram is defined by the region where, simultaneously, the (3,1) charge state is lower in energy than the (3,2) charge state and the (3,2) ground state energy is lower in energy than the sum of the (4,1) ground state energy and the (4,1) multi-electron orbital splitting. This region is shaded gray in Fig.~\ref{fig:fig1}(d). The gradient of the shaded region is representative of the multi-electron orbital splitting measured at varying electrostatic configurations shown in the inset. Using the screening gate S3, the orbital confinement of a quantum dot can be tuned \textit{in situ} as demonstrated in Ref.~\cite{McJunkin:2021p085406, Dodson:2021p14702}.

\begin{figure}[t!]
\includegraphics[width=0.45\textwidth]{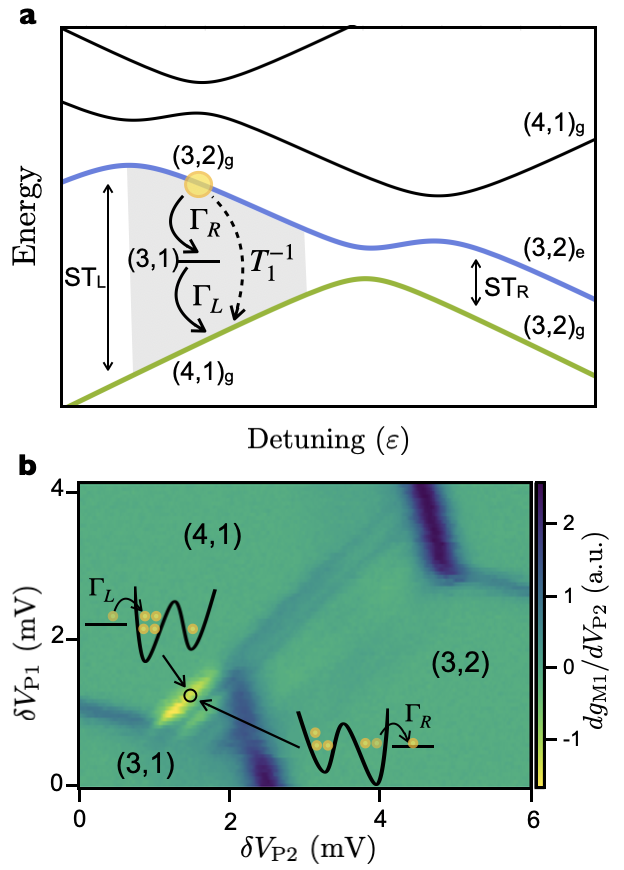}
\caption{\label{fig:fig2}  Latched readout of a 5-electron quantum dot hybrid qubit (QDHQ). (a) The energy eigenvalues of the qubit, where the logical $\ket{0}$ and $\ket{1}$ states are shown as the green and blue curves, respectively. Latched readout is possible in the gray region as long as the tunnel rate $\Gamma_R$ is faster than the decay rate from (3,2)$_g$ to (4,1)$_g$. When this condition is met, the (3,2) charge state is latched to the (3,1) charge state. (b) Measured response in the charge sensor with a lock-in tone and Larmor pulse applied to gate P2. The bright yellow response shows enhanced sensitivity in the latched readout regime. When the (3,2) charge state tunnels into the (3,1) configuration, the (3,1) charge state persists until the tunneling process $\Gamma_L$ occurs. $\Gamma_L$ can be tuned to be much longer than the decay rate from (3,2)$_g$ to (4,1)$_g$, enhancing the readout signal.  }
\end{figure}

Figure~\ref{fig:fig2} demonstrates latched readout of a 5-electron QDHQ. Fig.~\ref{fig:fig2}(a) shows the QDHQ energy eigenvalues, where the green and blue curves are the logical $\ket{0}$ and $\ket{1}$ states, respectively. The qubit is operated at positive double-dot-detuning energy $\varepsilon$, with the far-detuned qubit energy given by the singlet-triplet splitting in the right dot (ST$_R$). After ac or dc manipulation at positive $\varepsilon$, an adiabatic ramp changes $\varepsilon$ to the latched readout region, shown shaded in gray. If the qubit is in the logical $\ket{0}$ or $\ket{1}$ state, the resulting charge configuration in the readout regime will be the (4,1)$_\text{g}$ or (3,2)$_g$ charge state, respectively. In this way, the   logical basis is projected onto a charge basis that can easily be measured via the integrated charge sensing dot. Note that the decay from (3,2)$_\text{g}$ 

\noindent to (4,1)$_\text{g}$ does not require a spin-flip and is therefore expected to be fast, similar to a charge qubit \cite{Wang:2013p046801, Kim:2015p243,Scarlino:2019p206802,  MacQuarrie:2020p81}.

%The fast decay rate negatively impacts readout visibility of averaged measurements and makes single-shot readout unachievable \cite{Jang2021p00783} given current state of the art charge detection methods \cite{Curry:2015p203505, Tracy:2010p192110, Rossi:2017p84224, West:2019, Zheng:2019p04889, Liu:2021p14560}. 

By incorporating a latching mechanism, enhanced readout is achieved, as demonstrated in Fig.~\ref{fig:fig2}(b) using lock-in detection. In Fig.~\ref{fig:fig2}(b), the double dot is initialized in the (4,1)$_g$ state by waiting within the latching regime. Then, a Larmor pulse is applied, diabatically pulsing the detuning to the polarization line where the Hamiltonian suddenly changes, populating the logical $\ket{1}$ excited state. After the Larmor pulse duration, the detuning is pulsed back into the (4,1) charge configuration. The fringes both within and outside of the latched triangular region in Fig.~\ref{fig:fig2}(b) are indicative of Larmor oscillations resulting from the applied pulse. When the readout position lies within the latched readout regime, the charge sensed signal is enhanced \cite{Petersson:2010p246804,Studenikin:2012p233101}, as is evident by comparing the bright yellow fringes within the latched triangle to the blue ones outside the 

%width=1.0\textwidth
\onecolumngrid

\begin{figure}
\includegraphics[width=1.0\textwidth]{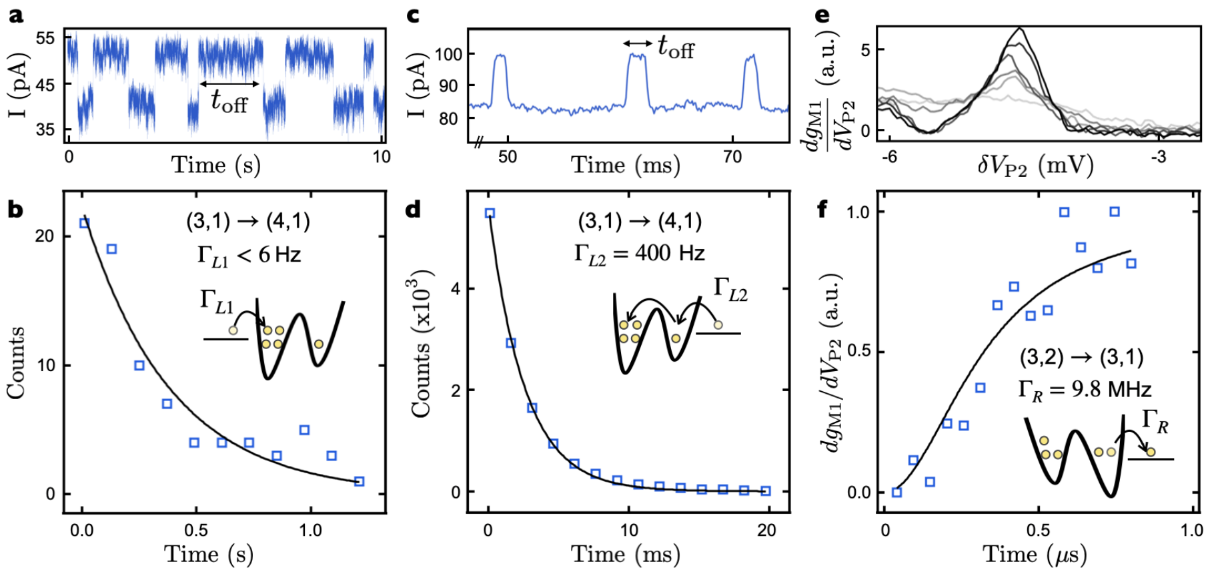}

\caption{ Measurement of tunnel rates $\Gamma_R$ and $\Gamma_L$. (a, b) The tunnel rate between the left reservoir and P1 dot ($\Gamma_{L1}$) is measured by real-time detection of single electron tunneling events and binning the amount of time the electron spends off the dot. The result is fit to the black line, yielding a tunnel rate of less than 6 Hz. (c, d) The same technique is used to measure the cotunneling rate from the right reservoir into the P1 dot ($\Gamma_{L2}$), where a faster tunnel rate of 400 Hz is measured. Thus, the tunnel rate from the left reservoir into the P1 dot is negligibly slow. (e, f) At the same electrostatic tuning, $\Gamma_R$ is measured to be 9.8 MHz. This is determined by applying a lock-in tone and a square pulse to gate P2 and measuring the peak height of the lock-in response as a function of the square pulse frequency. $\Gamma_R$ is purposefully made to be much faster than the other rates so that the tunneling event from (3,2)$_g$ to (3,1)$_g$ occurs before the (3,2)$_g$ to (4,1)$_g$ decay. The fast $\Gamma_R$ causes the cotunneling rate $\Gamma_{L2}$ to be the dominant process for the tunneling process $\Gamma_L$. }
\label{fig:fig3}
\end{figure}
\twocolumngrid

\noindent triangle in Fig.~\ref{fig:fig2}(b). Since the charge sensor is more sensitive to changes in total number of electrons, the capacitive shift for changes from (3,2) to (3,1) is larger than from (3,2) to (4,1).

In order for the latching mechanism to work successfully, the tunnel rate from (3,2)$_g$ to (3,1)$_g$, denoted as $\Gamma_R$ in Fig.~\ref{fig:fig2}, must be faster than the decay rate from (3,2)$_g$ to (4,1)$_g$ ($\Gamma_1 = T_1^{-1}$). Once this condition is met, the latching time can be tuned \textit{in situ} by changing the tunnel rate from (3,1)$_g$ to (4,1)$_g$, denoted as $\Gamma_L$ in Fig.~\ref{fig:fig2}. 

Fig.~\ref{fig:fig3} shows measurements of $\Gamma_L$ and $\Gamma_R$ at a single electrostatic tuning that is well suited for latched readout. For $\Gamma_L$ = $\Gamma_{L1} + \Gamma_{L2}$ there are two tunneling processes that contribute: the tunnel rate from the reservoir to the left of the P1 quantum dot ($\Gamma_{L1}$) and the cotunneling rate from the right reservoir ($\Gamma_{L2}$). Each of these tunnel rates are individually measured using real-time detection of single electron tunneling events in and out of the P1 quantum dot. A single time trace is taken where the position of the excess electron is mapped out in real time by measuring the current through the charge sensor. Fig.~\ref{fig:fig3}(a) shows a sample trace for 10 seconds where the higher (lower) current corresponds to the excess electron off (on) the dot. The amount of time spent off the dot between each tunneling event is binned and plotted in Fig.~\ref{fig:fig3}(b). The data is then fit using the results from Ref.~\cite{MacLean:2007p1499} and plotted as the black line. From the fit, a tunnel rate of $\Gamma_{L1} = 6$ Hz is extracted.  Note that this tunnel rate is much smaller than $\Gamma_{L2}$, whose measurement is described below, and $\Gamma_{L2}$ was temporarily tuned to a very low value while measuring $\Gamma_{L1}$.

A similar procedure is used for extracting $\Gamma_{L2}$, as shown in Fig.~\ref{fig:fig3}(c, d). Here, instead of direct tunneling, cotunneling is the dominant process, yielding a cotunneling rate of $\Gamma_{L2} = 400$ Hz. The direct tunneling rate from the left reservoir is purposefully made to be much slower than the cotunneling rate by lowering the voltage on gate B1. This maximizes the latching time,  which for this electrostatic tuning is 2.5 ms. 

The cotunneling rate is affected by both the interdot tunnel rate $t_c$ and the tunnel rate from P2 into the right reservoir ($\Gamma_R$) \cite{Nazarov:2009}. The double dot system is set at an electrostatic tuning that is realistic for QDHQ operation with a tunnel coupling of several GHz and a $\Gamma_R$ that is set to be fast such that it is comparable to the decay rate $\Gamma_1$. Fig.~\ref{fig:fig3}(e, f) shows the results of measuring $\Gamma_R$. A lock-in tone and square pulse at frequency $f_p$ are applied to gate P2 while measuring the amplitude of the lock-in signal as a function of the square pulse frequency. This procedure is discussed in greater detail in Ref.~\cite{Elzerman:2004p731}. The peak heights are normalized and plotted as blue squares in Fig.~\ref{fig:fig3}(f) as the pulse time $\tau_p = 1/2f_p$ is varied. The data is fit and shown as the black line, yielding $\Gamma_R = 9.8$ MHz. Given these values, single-shot readout of the QDHQ is possible in the future due to the long latching time of 2.5 ms.

\begin{figure}[t]
\includegraphics[width=0.50\textwidth]{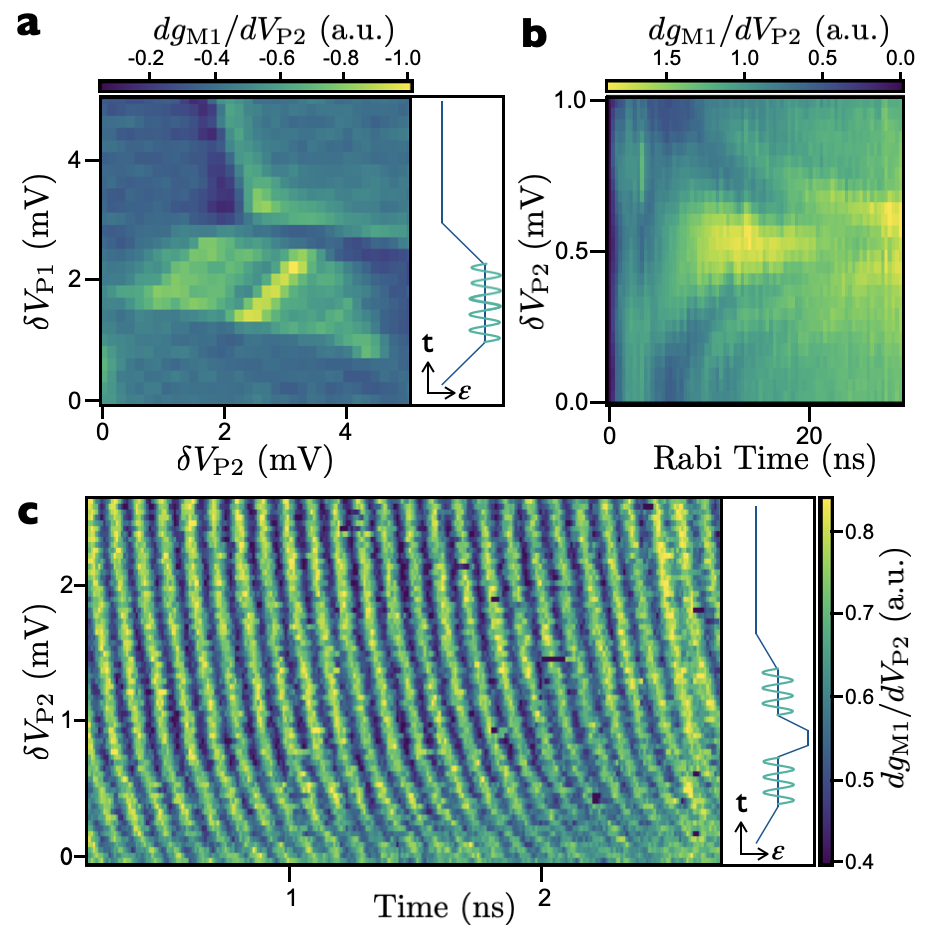}
\caption{\label{fig:fig4} Microwave-driven QDHQ using latched readout. (a) Latched readout zone with a Rabi pulse applied, as shown in the Inset. The applied pulse height and microwave drive are chosen to match the qubit energy at a location in positive detuning, and the latched signal appears as a bright line within the triangular region. (b,c) Rabi and detuned Ramsey oscillations driven with microwave frequency 10 GHz. The Inset in (c) shows the Ramsey pulse sequence where two $\pi/2$ microwave bursts are separated by a dc detuning pulse. }
\end{figure} 

Fig.~\ref{fig:fig4} shows resonant microwave control of the 5-electron QDHQ using lock-in measured latched readout. In Fig.~\ref{fig:fig4}(a), a stability diagram is measured with a Rabi pulse applied as pictured in the right-hand-side inset. With this pulse, the double quantum dot is initialized in the (4,1)$_g$ state and the detuning is adiabatically ramped into the (3,2)$_g$ charge regime. A resonant microwave burst is applied for a time $t_{\text{rabi}}$, populating the logical $\ket{1}$ state of the QDHQ. The detuning is adiabatically ramped back into the (4,1)$_g$ configuration, where the latching mechanism occurs by tuning the (3,2)$_g$ to (3,1)$_g$ tunnel rate to be faster than the (3,2)$_g$ to (4,1)$_g$ decay rate. When this condition is met, enhanced readout is possible, shown as the yellow triangular region in Fig.~\ref{fig:fig4}(a). The bright diagonal line within the triangle indicates the location in detuning where the applied microwave is resonant with the qubit energy.

Figure~\ref{fig:fig4}(b,c) shows Rabi and detuned Ramsey oscillations measured using latched readout. The Rabi measurement in Fig.~\ref{fig:fig4}(b) is performed by sweeping both the length of the microwave drive and the position of the pulse sequence within the latched window, controlled by $\delta V_{P2}$. The frequency of the microwave drive for both measurements was 10~GHz. 
%The extended coherence time as compared to charge qubits demonstrates that the 5-electron QDHQ is is less sensitive to charge noise at large detunings where the resonant microwave is applied. 
In Fig.~\ref{fig:fig4}(c), Ramsey fringes are observed in the presence of the pulse sequence shown in the inset. Here, two $\pi/2$ microwave bursts are separated by a dc detuning pulse, which allows for efficient measurement of the characteristic Larmor oscillation frequency as a function of detuning \cite{Thorgrimsson:2017p32, Corrigan:2021p127701}. The oscillations seen in Fig.~\ref{fig:fig4}(c) are measured as a function of the dc detuning pulse height and the wait time $t_{Ramsey}$ at the top of the pulse. The parallel oscillations as a function of increased $\delta V_{P2}$ are characteristic of the asymptotic qubit energy of a QDHQ at high detuning. 

In summary, here we identified and experimentally measured key parameters necessary for achieving latched readout of the QDHQ, and we demonstrated such latched readout of a 5-electron QDHQ. The tunnel rate from (3,2)$_g$ to (3,1)$_g$ is tuned to be faster than the (3,2)$_g$ to (4,1)$_g$ decay rate, and the direct tunnel rate from (3,1)$_g$ to (4,1)$_g$ is tuned to be slower than the cotunneling rate. These findings provide an important development for readout of QDHQs: the latched state persists much longer than that of a typical charge-mapped QDHQ excited state, providing a path for single-shot readout in the future. Additionally, cotunneling allows for latched readout of the QDHQ with only a single reservoir. Future implementations of QDHQs may use this readout architecture for multi-qubit operation and readout. 

We thank L.F.\ Edge for providing the Si/SiGe heterostructure used in this work. Research was sponsored in part by the Army Research Office (ARO) under Grant Number W911NF-17-1-0274. JC acknowledges support from the National Science Foundation Graduate Research Fellowship Program under Grant No. DGE-1747503 and the Graduate School and the Office of the Vice Chancellor for Research and Graduate Education at the University of Wisconsin-Madison with funding from the Wisconsin Alumni Research Foundation. We acknowledge the use of facilities supported by NSF through the UW-Madison MRSEC (DMR-1720415) and the NSF MRI program (DMR-1625348). The views and conclusions contained in this document are those of the authors and should not be interpreted as representing the official policies, either expressed or implied, of the Army Research Office (ARO), or the U.S. Government. The U.S. Government is authorized to reproduce and distribute reprints for Government purposes notwithstanding any copyright notation herein.

%\begin{acknowledgments}
%\end{acknowledgments}

\section*{Author Declarations}
\subsection*{Conflict of Interest}
The authors have no conflicts to disclose.
\subsection*{Author Contributions}
JC, JPD performed the measurements.  BT, TJK, TM contributed to the experimental setup and methods.  JP fabricated the sample with assistance from SFN.  JC, JPD, SNC, MAE analyzed the data.  All the authors wrote the manuscript.

\section*{Data Availability Statement}

The data reported here is available from the corresponding author on reasonable request.

\bibliography{main.bib}% Produces the bibliography via BibTeX.

\end{document}